\documentclass[a4paper,UKenglish]{lipics-v2018}

\usepackage{microtype}

\title{Fast Fibonacci heaps with worst case extensions}

\author{Vladan Majerech}{Department of Theoretical Computer Science and Mathematical Logic (KTIML), Charles University, Malostransk\'e n\'am\v est\'\i\ 25, Prague 118 00, Czech Republic}{maj@ktiml.mff.cuni.cz}{ https://orcid.org/0000-0003-3006-2002}{}

\authorrunning{V. Majerech}

\Copyright{Vladan Majerech}

\subjclass{Information systems$\rightarrow$Information storage systems$\rightarrow$Record storage systems$\rightarrow$Record storage alternatives$\rightarrow$Heap (data structure)}


\category{}

\relatedversion{}

\supplement{}

\funding{}

\acknowledgements{}

\EventEditors{}
\EventNoEds{1}
\EventLongTitle{}
\EventShortTitle{}
\EventAcronym{}
\EventYear{2019}
\EventDate{}
\EventLocation{}
\EventLogo{}
\SeriesVolume{}
\ArticleNo{1}
\nolinenumbers 
\hideLIPIcs  

\begin{document}
\newcommand{\prevfootnotemark}{\csname @footnotemark\endcsname}
\newcommand{\MakeHeap}{{\bf MakeHeap\/}}
\newcommand{\FindMin}{{\bf FindMin\/}}
\newcommand{\DeleteMin}{{\bf DeleteMin\/}}
\newcommand{\Delete}{{\bf Delete\/}}
\newcommand{\Meld}{{\bf Meld\/}}
\newcommand{\Cut}{{\bf Cut\/}}
\newcommand{\Insert}{{\bf Insert\/}}
\newcommand{\Decrement}{{\bf DecreaseKey\/}}
\newcommand{\Right}{\hbox{\sl right\/}}
\newcommand{\Left}{\hbox{\sl left\/}}
\newcommand{\nill}{\hbox{\sl null\/}}
\newcommand{\Parent}{{\sl parent\/}}
\newcommand{\numfootnote}{\footnote}

\maketitle

\begin{abstract}
We are concentrating on reducing overhead of heaps based on comparisons with optimal worstcase behaviour. 
The paper is inspired by Strict Fibonacci Heaps \cite{StrictHeaps}, where G. S. Brodal, G. Lagogiannis, and R. E. Tarjan implemented the heap with \Decrement\ and \Meld\ interface in assymptotically optimal worst case times (based on key comparisons). 
In the paper \cite{sewc}, the ideas were elaborated and it was shown that the same asymptotical times could be achieved with a strategy loosing much less information from previous comparisons. There is big overhead with maintainance of violation lists in these heaps. 
We propose simple alternative reducing this overhead. 
It allows us to implement fast amortized Fibonacci heaps, where user could call some methods in variants guaranting worst case time.
If he does so, the heaps are not guaranted to be Fibonacci until an amortized version of a method is called.
Of course we could call worst case versions all the time, but as there is an overhead with the guarantee, calling amortized versions is prefered choice if we are not concentrated on complexity of the separate operation.

We have shown, we could implement full \Decrement-\Meld\ interface, but \Meld\ interface is not natural for these heaps, so if \Meld\ is not needed, much simpler implementation suffices. As I don't know application requiring \Meld, we would concentrate on no\Meld\ variant, but we will show the changes could be applied on \Meld\ including variant as well. The papers \cite{StrictHeaps}, \cite{sewc} shown the heaps could be implemented on pointer machine model.
For fast practical implementations we would rather use arrays. Our goal is to reduce number of pointer manipulations. Maintainance of ranks by pointers to rank lists would be unnecessary overhead.
\end{abstract}

\section{Introduction}
We will call heaps from the paper \cite{StrictHeaps} BLT heaps as their connection to Fibonacci is only negligable.
We will call the heaps in paper \cite{sewc} as BLMT heaps, with \Meld\ and no\Meld\ variants. 
The BLT (resp. BLMT) heaps strategy is not to be too pedantic to the heap shape and allow some types of violations.
Violation sizes of each type are maintained bounded by maximal rank $R$ plus 1. 
What gives quadratic inequality bound for $R$ leading to $R\le 6+1.2\log_2 n$. To maintain degrees bounded by $O(\log n)$ in \Meld\ variants, list of heap nodes is required and 
first two nodes are moved to the end of the list after each heap size decrement by 1. The degree reduction is performed on moved nodes.
This ensures the $2n-p$ ($p$ position in the list) remains constant for all except moved nodes so degree bounds by funcion $b(2n-p)$ could be maintained.
For no\Meld\ variant no degree reductions are required and no list of heap nodes is needed.

The violations are maintained in violation structures in BLT/BLMT heaps. 
In conlcluding remarks of \cite{sewc} the big overhead with their maintainance is mentioned and caching is recomended. 
Our goal is to reduce violation structures at all, and use same rank identifying places with caching instead.
Amortized versions would simply empty the caches at the end of each operation. 
Worst case versions would need to plan the reductions and count the reductions made not to exceed the declared time. 
We have to propose violation/cache reduction strategy to ensure the violation sizes remain bounded by $R+1$ at the same time.

With empty cache for loss violations we have guarentee no node have loss greater 1, what is condition maintained at Fibonacci heaps, therefore 
at these times Fibonacci sequence bounds the sizes of subtrees of nodes of given rank.
Actually for big $n$ the total loss bound by $R(n)$ is more restrictive than local loss bound at each node. 
So only for small ranks (and empty cache) the Fibonacci bound applies. The trees with bigger ranks are bounded more by total loss.
This is on the edge to call them Fibonacci. Calling them (local/)global loss bounded would be more appropriate.

\section{Overhead of violation structures}
Violations in BLT/BLMT heaps are maintained in double linked lists pointed inside from ranks. 
Adding a violation first of the rank means making pointer from rank to the node and inserting listnode to the corresponding list end. 
This is why end list pointer changes and 2 pointers in neighbours are updated (and one remains \nill) so at least 4 pointer updates are required.
Adding 2nd violation of the same rank removes node pointed by rank from the list, and inserts pair of nodes to the other end of the list.
This changes 2 neighbour pointers around removed node from the list, sets 2 pointers among the pair of nodes of the same rank and changes end list pointer and 2 pointers connecting the inserted pair with the list. This makes 7 pointer changes.
When node with loss 1 gets 2nd loss, it was pointed by rank pointer and there is exactly one other node of the same rank in the violation list, 
the update is even bigger. Rank pointer should be changed, the other node of the rank should be removed from the list and reinserted to the end of the list.
The loss 2 node representant should be removed from the list as well and reinserted to the other end of the list.
This affects 2 pointers to reconnect the list near removal, both list end pointers change and both moved nodes change both neighbours (\nill\ at end of the list should be changed to inserted node, inserted node should change one pointer to neighbour and the other to \nill).
This sums to 11 pointer changes.
Violation reduction steps remove upto two nodes from corresponding list end and either update rank pointer or if there is just one node remaining of affected rank, the node is removed and reinserted to the other list end. This affects at most 6 pointers (not counting overhead by garbage collector of reduced nodes).

We propose using same rank identifying places and cache of not yet inserted violations. On pointer machine model the list of ranks could be used as same rank identifying places (with one pointer reserved in the list node for each type of violation). 
We would recomend array for each type of violation indexed by numeric ranks when pointer machine model is not required.

Cache could be any set structure supporting insert of a pointer value and pop (removing an arbitrary pointer value from the set and returning it).
Single link list could be used on pointer machine requiring 2 pointer changes per update. Array (with guaranted sufficient size) requiring 1 pointer change and one index value change per insert and no pointer change and one index value change per pop is recomended when pointer machine model is not needed.

Each violation would be inserted to the cache and cache would be processed during cache reductions. 
As each node could be part of at most one violation type, we would maintain the type identification in the node (could be empty, loss $\ge 2$ could be identified here as well).
When node changes rank, it allows us to localise the violation set to update the node rank here (to remove, change, and reinsert).
The removal from corresponding type of violation would check corresponding pointer in same rank identifying place. 
If it points to the node, we just change it to \nill. 
Otherwise we just remember it is in the cache. 
This would leave the violation update in cache. 
If the node should be reinserted to the same violation type we insert it to cache (unless we know it is there already).
If the node should be reinserted as other violation type, we insert it to the corresponding cache, but update the type of violation to which it should belong.
If the node is no more violation, we should empty the type identification maintained in it.
This means node could be even several times in several caches, but it would represent at most one violation of its current violation type.

When node from cache is processed during cache reduction, we at first check it belongs to the type of violation. If not, we just discard the cache info.
We will discard the cache info even when corresponding rank identifying place points to the node.
When the node is loss $\ge 2$ violation (allowed on loss violation cache), we do corresponding single node loss reduction immediately. 
Otherwise we check the same rank identifying place for the node rank. 
If it contains pointer to the processed node, we just discard the cache info.
If it contains \nill, the pointer to the node is inserted there.
Otherwise we have pointers to two nodes of the same rank and corresponding violation reduction is performed. This puts \nill\ to the same rank identifying place and could insert new violations to corresponding caches.

We actually use same rank identifying places the way it is used in \FindMin\ of amortised versions of Binomial/Fibonacci/Padovan heaps except 
we leave the pointers in the places even after \FindMin\ is finished. This reduces the overhead with same rank identifying places of Binomial/Fibonacci/Padovan heaps.

The arrays would be preferable choice regarding to hardware caches.
Using arrays in no\Meld\ variant is natural, especially when we could predict the maximal heap size, otherwise we could use worstcase variant of array doubling. The array doubling overhead would be negligable as the arrays are of logarithmic sizes. 
Worstcase variant of array doubling would work even in \Meld\ variant, it would need to copy two slots per array and operation in scenario long sequence of melds of equally sized heaps occurs. Most other times the array would be filled less than from half so no copying is needed.

\section{Structure balancing overview}
Except the introduction of caches and change on their maintanance strategy the heaps would work as in \cite{sewc}.
In the \Meld\ variant there would be deffered and solid nodes, where deffering could be implicit.
Solid children would be either rank or nonrank. The number of rank children correspond to rank of the node which is bounded by some function $R(n)$.
Solid nonrank child is a rank tree root. We will maintain all rank tree roots as violations to ensure, their number does not exceed $R(n)+1$.
Unfortunately if we would maintain all rank tree roots together, violation reductions would allow series where 
increase of degree would force degree reduction resulting in new rank tree root violation so there will be no visible progress in reducing rank tree roots count. Therefore we crerate two violation types for rank tree roots, one where the roots with guaranted degree reserves are added and the others (this prevents eager conversion of most defered children to solid).
Each degree reduction ensures the guarantee and linking two guaranted rank roots does not require the degree reduction.
Therefore the number of rank roots is bounded by $2R(n)+2$, one of them is the main root so at most $2R(n)+1$ rank roots could be children.

We would maintain all nodes with loss as violation of loss type.  
The number of solid children is therefore bounded by $3R(n)+1$ where $R(n)$ is maximal posible rank for heap with $n$ nodes and total loss at most $R(n)+1$.
Estimating the bigger root of the corresponding quadratic equation gives us $R(n)\le 6+1.2\log_2 n$.

In the case there could be deffered children (\Meld\ variant) we should maintain number of children bounded by $O(\log n)$ explicitly.
It is sufficient to regularly do node degree reductions. 
This reduction either reduces the degree of a node by 2 or ensures the node has no more than 2 deffered children.
Each \Meld\ ensures the bounds hold for newly implicitly deffered nodes and the bounds do not decrease for other nodes.
Whenever heap size is decremented (due to \DeleteMin) two heap nodes are degree reduced and their bounds are changed.
List of heap nodes, removing first two nodes from it and inserting them as last two (after two node degree reductions) would do needed, 
when bounds are defined by function of $3R(2n-p)+c$, where $p$ is position of the node in the list, and $c$ is small natural number depending on node being solid without loss or not. Actually $c$ allows one more children (4+1) for solid nodes with zero loss than for others (4).
The degree reductions would maintain the bounds implicitly without knowing their actual values.

We would present the \Meld\ variant first and the simple no\Meld\ variant at the end.

\section{Violation reductions}
\vbox to 86mm{
\hbox{\kern7mm\pdfximage width 14cm {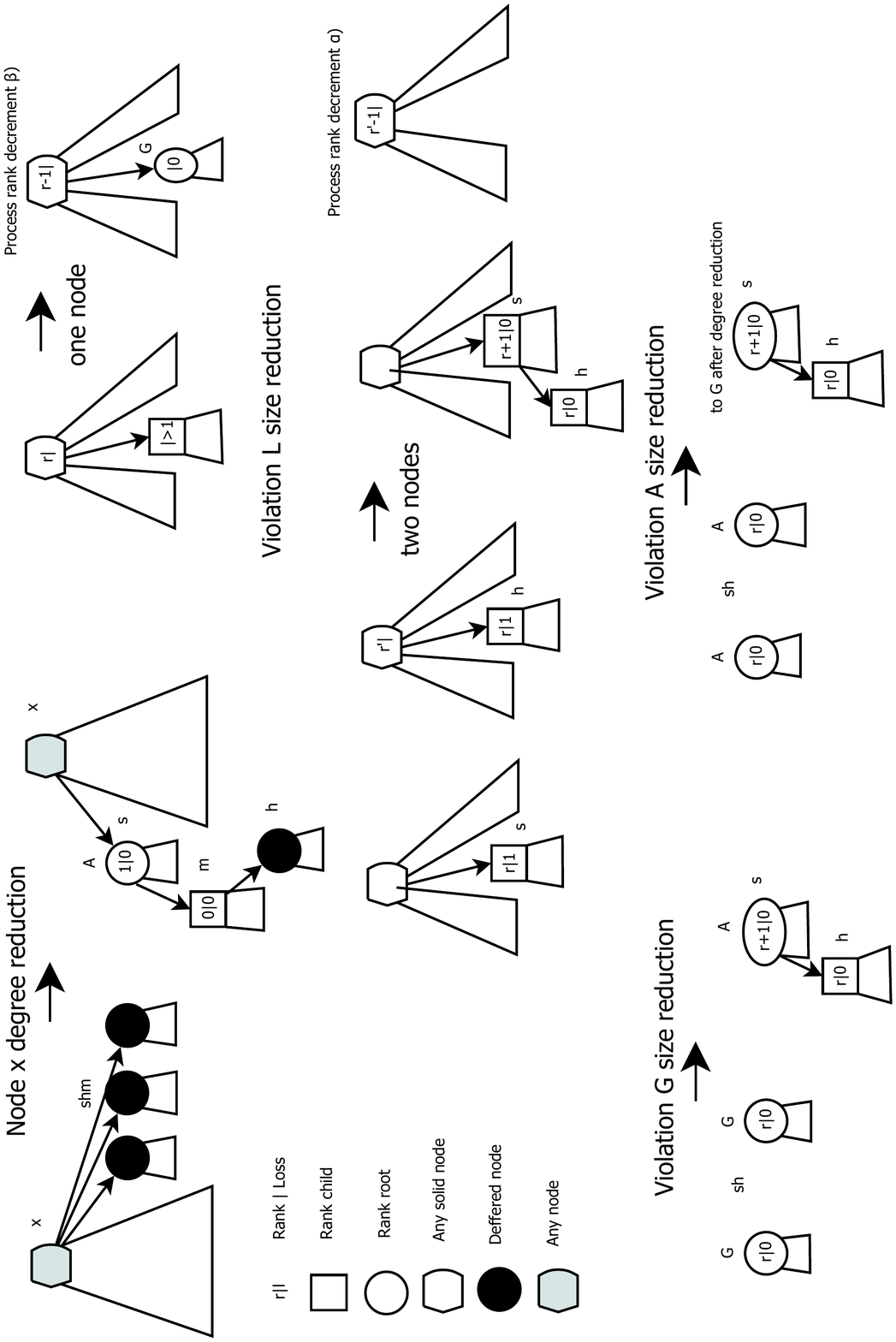}
\rlap{\smash{\pdfsave\pdfsetmatrix{0 -0.56 0.56 0}
\pdfrefximage\pdflastximage}}\pdfrestore
\hss}
\vss
\hbox{Figure 1: Reductions to maintain the heap shape with complications for \Meld\ support}
\kern4mm
}

\begin{table}
 \begin{center}
  \caption{Effect of different transformations $3|GR|+4|GC|+5|AR|+6|AC|+10|LR|+11|LC|=\Phi$}
	\label{tab:eff}
	\begin{tabular}{lrrrrrrrr}
	 Reduction & $|GR|$ & $|GC|$ & $|AR|$ & $|AC|$ & $|LR|$ & $|LC|$ & $\Phi$ & $p$\\
	 \hline
	 \hline
	 node degree & $0$ & $0$ &  $0$ & $+1$ & $0$ & $0$ & $+6$ & $1$\\
	 \hline
	 $|AC|$ discard & $0$ & $0$ & $0$ & $-1$ & $0$ & $0$ & $-6$ & $0$ \\
	 \hline
	 $|AC|$ type $A$ no match & $0$ & $0$ & $+1$ & $-1$ & $0$ & $0$ & $-1$ & $1$ \\
	 \hline
	 $|AC|$ type $A$ matched & $0$ & $+1$ & $-1$ & $\le 0$ & $0$ & $0$ & $\le -1$ & $\le 3$\\
	 \ - no 3 deffered chidren & $0$ & $+1$ & $-1$ & $-1$ & $0$ & $0$ & $-7$ & $2$ \\
	 \ - 3 deffered children & $0$ & $+1$ & $-1$ & $0$ & $0$ & $0$ & $-1$ & $3$ \\
	 \hline
	 $|GC|$ discard & $0$ & $-1$ & $0$ & $0$ & $0$ & $0$ & $-4$& $0$ \\
	 \hline
	 $|GC|$ type $G$ no match & $+1$ & $-1$ & $0$ & $0$ & $0$ & $0$ & $-1$& $1$ \\
	 \hline
	 $|GC|$ type $G$ matched & $-1$ & $-1$ & $0$ & $+1$ & $0$ & $0$ & $-1$ & $2$\\
	 \hline
	 $|LC|$ discard & $0$ & $0$ & $0$ & $0$ & $0$ & $-1$& $-11$ & $0$ \\
	 \hline
	 $|LC|$ subtype $L'$ & $\le 0$ & $\le +2$ & $\le 0$ & $\le +1$ & $\le 0$ & $\le 0$ & $\le -1$ & $\le 3$\\
	 \ - parent G in $GR$& $-1$ & $+2$ & $0$ & $0$ & $0$ & $\le -2$ & $\le -17$ & $3$ \\
	 \ - parent G in $GC$& $0$ & $+1$ & $0$ & $0$ & $0$ & $\le -2$ & $\le -18$ & $1$ \\
	 \ - parent A in $AR$& $0$ & $+1$ & $-1$ & $+1$ & $0$ & $\le -2$ & $\le -17$ & $3$ \\
	 \ - parent A in $AC$& $0$ & $+1$ & $0$ & $0$ & $0$ & $\le -2$ & $\le -18$ & $1$ \\
	 \ - parent L in $LR$& $0$ & $+1$ & $0$ & $0$ & $-1$ & $\le 0$ & $\le -6$ & $3$ \\
	 \ - parent L* in $LC$& $0$ & $+1$ & $0$ & $0$ & $0$ & $\le -1$ & $\le -7$ & $1$ \\
	 \ - parent N no 3 deffered children& $0$ & $+1$ & $0$ & $0$ & $0$ & $\le -1$ & $\le -7$ & $2$ \\
	 \ - parent N 3 deffered children& $0$ & $+1$ & $0$ & $+1$ & $0$ & $\le -1$ & $\le -1$ & $3$ \\
	 \hline
	 $|LC|$ subtype $L$ no match & $0$ & $0$ & $0$ & $0$ & $+1$ & $-1$& $-1$ & $1$ \\
	 \hline
	 $|LC|$ subtype $L$ matched & $\le 0$ & $\le +1$ & $\le 0$ & $0$ & $\le 0$ & $\le +1$ & $\le -9$ & $\le 3$ \\
	 \ - parent of $h$ G in $GR$& $-1$ & $+1$ & $0$ & $0$ & $-1$ & $-1$ & $-20$ & $3$ \\
	 \ - parent of $h$ G in $GC$& $0$ & $0$ & $0$ & $0$ & $-1$ & $-1$& $-21$ & $1$ \\
	 \ - parent of $h$ A in $AR$& $0$ & $+1$ & $-1$ & $0$ & $-1$ & $-1$ & $-22$ & $3$ \\
	 \ - parent of $h$ A in $AC$& $0$ & $+1$ & $0$ & $0$ & $-1$ & $-1$ & $-17$ & $2$ \\
	 \ - parent of $h$ L in $LR$& $0$ & $0$ & $0$ & $0$ &$-2$ & $+1$ & $-9$ & $3$ \\
	 \ - parent of $h$ L* in $LC$& $0$ & $0$ & $0$ & $0$ & $-1$ & $0$ & $-10$ & $1$ \\
	 \ - parent of $h$ N& $0$ & $0$ & $0$ & $0$ & $-1$ & $0$ & $-10$ & $2$ \\
	 \hline
	\end{tabular}
 \end{center}
 \vskip 4pt
 Here $p$ denotes number of pointer changes not reflected in heap trees during reduction when arrays are used for caches.
 We can see each cache reduction decrements $\Phi$ by at least 1.
\end{table}

Deffered nodes would be made by the \Meld\ method.
Similarly as in BLT heaps, the nodes of smaller heap would become deffered implicitly.
Implicitly deffered nodes cannot have solid children.
Deffered nodes would be accessed during degree reductions of their parents, and during \DeleteMin s, 
when the heap root was deffered node parent or by degree reduction of the deffered node reflecting decrease of heap size when moved from the start of heap node list to its end. When the implicitly deffered node is firstly accessed, the pointer responsible for implicit deffering is removed and the node is converted to explicitly deffered. Its heap pointer is redirected to current heap. All its children are deffered (either implicitly or explicitly) in this time, so they are rightmost. No solid rank child is allowed for explicitly deffered nodes, but nonrank solid children are allowed.
Deffered nodes violation type is $N$ (no violation) by default.

Degree reduction step on a node $x$ would be made similarly as root degree reduction on BLT heaps.
If node $x$ is implicitly deffered, it is converted to explicitly deffered.
If the rightmost 3 children are deffered, we convert them to explicitly deffered if not converted yet and we remove them from children list of $x$.
We made 3 comparisons to find order of their keys, let node $s$ have the smallest, $m$ the middle, and $h$ the highest key.
We continue by making $s$ and $m$ solid.
We create rank edge making $s$ root of rank 1 having solid rank child $m$ of rank 0.
We make $h$ a deffered child of $m$, whose rank would stay 0. 
Finaly $s$ is linked as a nonrank (leftmost) child of $x$.
Degree constraints are OK for $s$ and $m$ as they become solid with loss 0.
New rank root without guaranted degree reserve was created. Degree of $x$ was reduced by 2.

Rank roots without guaranted reserve would be maintained as violations of type $A$. 
Violations of type $A$ would be inserted to cache $AC$ and from the cache to the same rank identifying places $AR$.
Similarly rank roots with guaranted reserve would be maintained as violations of type $G$ using cache $GC$ and same rank identifying places $GR$.

During $AC$ reduction if processed cache node $x$ is already not of type $A$, the cache item is discarded.
The cache item is discarded as well if $AR$ for $x$'s rank points to $x$.
Other case is $AR$ for $x$'s rank contains \nill, than pointer $x$ is stored there, and $AC$ reduction step ends.
Last, and the most important case is when $AR$ for $x$'s rank pointed to other violation $y$ of the same rank and actual violation type $A$, reduction step could be applied after putting \nill\ to $AR$ of $x$'s rank. 
Violation reduction step of type $A$ links nodes $x$, $y$ of the same rank. 
(Their keys are compared, let node $s$ be the one with smaller key while $h$ the other. We cut $h$ from its parent (if there exists nonrank edge) and put it as a rank child of $s$. This increases rank of $s$ as well as it's degree. Degree reduction is performed on $s$ what makes $s$ active root with guaranted reserve. So $h$ violation type is changed to $N$, the active root possibly created by the degree reduction would be added to $AC$. Node $s$ violation type is changed to $G$, and $s$ is added to $GC$.)

During $GC$ reduction the simillar trivial cases appear (just use $G$, $GC$ resp. $GR$ instead of $A$, $AC$ resp. $AR$).
Last, and the most important case is violation reduction step of type $G$ linking nodes $x$, $y$ 
of the same rank after putting \nill\ to $GR$ of $x$'s rank.
(Their keys are compared, let node $s$ be the one with smaller key while $h$ the other. We cut $h$ from its parent (if there exists nonrank edge) and put it as a rank child of $s$. This increases rank of $s$ as well as it's degree. Degree reduction is not performed on $s$ as there was degree reserve. So $h$ violation type is changed to $N$, $s$ violation type to $A$ and $s$ is added to $AC$.)

Whenever rank child's rank is decremented, its loss is incremented.
All nodes with nonzero loss would be maintained as violations of type $L*$, 
this type has subtype $L$ for nodes with loss exactly 1 and subtype $L'$ for nodes with loss at least 2.
Violations of type $L*$ would be inserted to cache $LC$, from which violations of subtype $L$ will be inserted to same rank identifying places $LR$.
Symbol $|LC|$ has weighted meaning. Weight of nodes of subtype $L'$ corresponds to the loss of the node, while others weight is 1 (including nodes of other type than $L*$).

Similarly as for $AC$ reduction, when during $LC$ reduction node $x$ is already not of type $L*$ or $LR$ of $x$'s rank points to $x$, the cache item is discarded.
Different is the second case when $x$'s subtype is $L'$. 
It invokes one node loss reduction, which takes node $x$ with loss at least 2, it makes it nonrank child of it's parent $p$.
This creates new rank root $x$ (with loss 0 and guaranted degree reserve), so $x$ is put to $GC$ and violation type of $x$ is changed to $G$.
The rank of $p$ is decremented.
Unless violation type of $p$ is $N$, $p$ should be removed from the rank identifying place identified by its type (If there was \nill\ in the place, we know $p$ resists in the cache).
If $p$ is a rank child it should be inserted to $LC$ and type changed to $L*$ (if it does not resist there), its loss is increased and subtype changed accordingly. Total loss was reduced by at least 1.
Degree of $p$ could have been on it's limit and the limit was decremented if the loss changed from 0 to 1, therefore degree reduction should be called on $p$ if it changed loss from 0 to 1 (what could insert new violation of type $A$ to $AC$).
If $p$ was a rank root, its violation type does not change as both limit and degree did not changed so $p$ should be just inserted to the cache of the type unless it already resists there.
Third case of $LC$ reduction is for $L$ subtype when the rank identifying place $LR$ of $x$'s rank contains \nill.
As for $AC$ reduction the pointer to $x$ is stored in $LR$ and the $LC$ reduction step ends.
Last case is when $LR$ for $x$'s rank (for node of subtype $L$) pointed to other violation $y$ of the same rank and actual violation type $L$ reduction step could be applied after putting \nill\ to $LR$ of $x$'s rank. 
Violation reduction step of type $L$ for nodes $x$, $y$ of equal rank and loss 1 links the two nodes. 
(Their keys are compared, let $h$ and $s$ be the nodes with higher and smaller keys respectively. 
 Remove $h$ from it's parent and link it under $s$ by a rank edge. 
 This reduces loss of $s$ to 0 and sets loss of $h$ to 0, so both $s$ and $h$ violation types are changed to $N$.
 Original parent $p$ of $h$ decrements rank by 1. 
 Unless violation type of $p$ is $N$, 
 $p$ should be removed from the rank identifying place identified by its type (If there was \nill\ in the place, we know $p$ resists in the cache).
 If $p$ is a rank child its type should be changed to $L*$ and $p$ inserted to $LC$ (if it does not resist there),
 and its loss is increased and subtype changed  accordingly. Total loss was reduced by at least 1.
 Degree constraint for $s$ is OK as well as for $p$.
 If $p$ was rank root, it got degree reserve so its type should be changed to $G$ and $p$ should be inserted to $GC$ (if it does not resist there).)

In the Figure 1 you can see the reductions and in Table \ref{tab:eff} you can see the effect of reductions.

For amortized versions, we do cache reductions unless all caches are empty. 
The process must terminate as $\Phi=3|GR|+4|GC|+5|AR|+6|AC|+10|LR|+11|LC|$ is decremented by each cache reduction. $\Phi$ could be used as potential to pay for the cache reductions.
Strategy to maintain violations in bounds would for worst case variants calculate changes of $\Phi_G=3|GR|+4|GC|$, $\Phi_A=5|AR|+6|AC|$, $\Phi_L=10|LR|+11|LC|$ from the start of the method. 
In each coordinate the positive change at the method end is allowed only in the case the corresponding cache is empty.
So while a coordinate has positive change and nonempty cache, the corresponding cache reduction step is invoked.
Again as $\Phi=\Phi_G+\Phi_A+\Phi_L$ is decremented by each cache reduction, we could simply bound the number of required reductions.

Let $\Delta^0 \Phi_G$, $\Delta^0 \Phi_A$, $\Delta^0 \Phi_L$ be the values at the start of reducing process, let 
$\Delta^1 \Phi_G=\max(-6,\Delta^0 \Phi_G)$, $\Delta^1 \Phi_A=\max(-4,\Delta^0 \Phi_A)$, and $\Delta^1 \Phi_L=\max(-10,\Delta^0 \Phi_G)$.
Violation reductions could not finish earlier providing we start with at least same coordinates of $\Delta\Phi$ so if we bound the number of violation steps providing we start at $\Delta^1 \Phi_G$, $\Delta^1 \Phi_A$, $\Delta^1 \Phi_L$, this would bound the real value.

Let $\Delta^E \Phi_G$, $\Delta^E \Phi_A$, $\Delta^E \Phi_L$ be the values at the end of reducing process starting from $\Delta^1 \Phi_G$, $\Delta^1 \Phi_A$, $\Delta^1 \Phi_L$.
3 deffered children case reduces $\Phi$ coordinates at least as no 3 deffered children case, so we can exclude it from analysis, as well as cases when cache item is just discarded. 
Let us exclude $|LC|$ reducing case which decrement $\Sigma_A$ because it reduces $\Phi$ coordinates at least as the next case in the table.
Now all remaining cases reduce only one coordinate, remaining coordinates could be only increased. 
Than $\Delta^E \Phi_G\ge -6$ as each reduction of $|GC|$ decreases $\Phi_G$ by at most 7.
Similarly $\Delta^E \Phi_A\ge -4$ as each nonexcluded reduction of $|AC|$ decreases $\Phi_A$ by at most 5.
If we consider the last |LC| decreasing step as decreasing $|LR|+|LC|$ by 1, change of $\Phi$ would be still at most $-1$ and $\Phi_L$ would change by at most $-11$. As $\Phi$ decreases by each considered reduction by at least $-1$, there could be at most $\Delta^1\Phi_G-\Delta^E \Phi_G+\Delta^1 \Phi_A-\Delta^E \Phi_A+\Delta^1 \Phi_L+10\le 
\max(-6,\Delta^0\Phi_G)+\max(-4,\Delta^0 \Phi_A)+\max(-10,\Delta^0 \Phi_L)+20$ reduction steps in total.

We know $|GR|\le R(n)+1$, $|AR|\le R(n)+1$, and $|LR|\le R(n)+1$. This defines equilibrium values of $\Phi_G\le 3R(n)+3$ for the case $|GC|=0$, $\Phi_A\le 5R(n)+5$ for the case $|AC|=0$, and $\Phi_L\le 10R(n)+10$ for the case $|LC|=0$. We will always use amortized versions of \DeleteMin\ which has the worstcase time $O(\log n)$.
This guarantees after each heap size decrement $\Phi$ coordinates would be at at most equilibrium values. 
When a worstcase method is called, we got either smaller $\Phi$ coordinate or empty cache so a $\Phi$ coordinate never increases above the equilibrium value. But the $\Phi_G$ bounds $|GR|+|GC|$ by $\Phi_G/3$, $\Phi_A$ bounds $|AR|+|AC|$ by $\Phi_A/5$ and $\Phi_L$ bounds $|LR|+|LC|$ by $\Phi_L/10$ so number of nodes of violation type $G$ is bounded by $|GR|+|GC|\le \Phi_G/3\le R(n)+1$, number of nodes of violation type $A$ is bounded by $|AR|+|AC|\le \Phi_A/5\le R(n)+1$, and total loss is bounded by $|LR|+|LC|\le \Phi_L/10\le R(n)+1$.

Degree reduction gives us equilibrium bounds for node degrees. We already know the number of solid children does not exceed $3R(n)+1$.
If it has at least $3R(n)+4$ children, at least 3 must be deffered and degree reduction can be performed. 
For our analysis it would be fine to define degree bounds $b(2n-p)=3R(2n-p)+5$ for solid nodes with loss 0 and $b(2n-p)=3R(2n-p)+4$ for other nodes (with $p$ being position in the global list of nodes).
With estimate $R(2n-p)\le 6+1.2\log_2(2n-p)$ we got $b(2n-p)\le 23+4\log_2(2n-p)$ resp. $b(2n-p)\le 22+4\log_2(2n-p)$ so we have to plan degree reduction by 4 when moving from start of the heap node list to its end to compensate for $n$ decrement, what corresponds to planning two degree reduction steps for a checked node.

As in BLT heaps, linking of rank roots which are nonrank children introduces situation which cannot happen when comparing only tree roots.
In the case keys could be equal, random choice of result would allow chosing $h$ to be predecessor of $s$ resulting in broken tree and a cycle.
To prevent this we should expect keys are all different. 
If this is not guaranted from outside, solution is to generate (different) id's for key nodes and broke ties by id's comparisons.

\section{Heap structure}

Heap information contains size info inicialized to 0, reference count initialised to 0, pointers to list of heap tree roots, and to the list of all heap nodes. 
It contains same rank identifying places $GR$, $AR$, $LR$ and the caches $GC$, $AC$ and $LC$.
All the lists are maintained double linked, left pointers are maintaned cyclic (left of leftmost points to rightmost). 
This allows access of both ends in constant time as well as adding or removing of a given node. 
List of heap tree roots uses sibling pointers maintained in the heap nodes. All heap nodes could have pointers internally in heap nodes as well.

In the pointer machine case we would maintain double linked list of ranks and rank would be represented by poiter to it. 
In the array version this is not required and we use arrays with worstcase doubling instead.
We would implement caches as stacks (last in first out).

When the size info is $-1$ and the reference count is 0, the heap information is discarded.

Whether the node is rank child, nonrank child or explicitly deffered is maintained in the node state, but this is overriden by being a heap tree root or being implicitly deffered.

Each node which points to the heap information with size info $<0$ is implicitly deffered. 
It could be made explicitly deffered by pointing to current heap and setting corresponding state. The reference counter for original heap should be decremented and reference counter on current heap incremented. 

\section{Implementation of methods}

We will describe the methods using private blocks. Their use could be slightly optimized (for example replacing pointer twice during a method
without reading it between changes could be avoided). Decomposition into blocks makes the description easier.

\begin{table}
 \begin{center}
  \caption{Effect of private blocks to violations}
	\label{tab:privblock}
	\begin{tabular}{lrrrrr}
	 Method & $\Phi_G$ & $\Phi_A$ & $\Phi_L$ & $\Phi$ & $p$\\
	 \hline
	 \hline
   heap size decrement & $0$ & $\le 24$ & $0$ & $\le 24$ & $\le 7$ \\
	 set violation type $G$& $\le +4$ & $\le 0$ & $\le 0$ & $\le +4$ & $\le 2$ \\
	 \ | from $G$ & $\le +1$ & $0$ & $0$ & $\le +1$ & $\le 2$ \\
	 \ | from $A$ & $+4$ & $\le 0$ & $0$ & $\le +4$ & $\le 2$ \\
	 \ | from $L*$ & $+4$ & $0$ & $\le 0$ & $\le +4$ & $\le 2$ \\
	 \ | from $N$ & $+4$ & $0$ & $0$ & $+4$ & $1$ \\
	 set violation type $A$& $\le 0$ & $\le +6$ & $\le 0$ & $\le +6$ & $\le 2$ \\
	 \ | from $G$ & $\le 0$ & $+6$ & $0$ & $\le +6$ & $\le 2$ \\
	 \ | from $A$ & $0$ & $\le +1$ & $0$ & $\le +1$ & $\le 2$ \\
	 \ | from $L*$ & $0$ & $+6$ & $\le 0$ & $\le +6$ & $\le 2$ \\
	 \ | from $N$ & $0$ & $+6$ & $0$ & $+6$ & $1$ \\
	 set violation type $L*$& $\le 0$ & $\le 0$ & $\le +12$ & $\le +12$ & $\le 2$ \\
	 \ | from $G$ & $\le 0$ & $0$ & $+11$ & $\le +11$ & $\le 2$ \\
	 \ | from $A$ & $0$ & $\le 0$ & $+11$ & $\le +11$ & $\le 2$ \\
	 \ | from $L*$ & $0$ & $0$ & $\le +12$ & $\le +12$ & $\le 2$ \\
	 \ | from $N$ & $0$ & $0$ & $+11$ & $+11$ & $1$ \\
	 set violation type $N$& $\le 0$ & $\le 0$ & $\le 0$ & $\le 0$ & $\le 1$ \\
	 \ | from $G$ & $\le 0$ & $0$ & $0$ & $\le 0$ & $\le 1$ \\
	 \ | from $A$ & $0$ & $\le 0$ & $0$ & $\le 0$ & $\le 1$ \\
	 \ | from $L*$ & $0$ & $0$ & $\le 0$ & $\le 0$ & $\le 1$ \\
	 \ | from $N$ & $0$ & $0$ & $0$ & $0$ & $0$ \\
	 rank decrement& $\le +4$ & $\le +1$ & $\le +12$ & $\le +12$ & $\le 2$ \\
	 \ | $\alpha$ rank root & $\le +4$ & $\le 0$ & $0$ & $\le +4$ & $\le 2$ \\
	 \ | $\beta$ rank root & $\le +1$ & $\le +1$ & $0$ & $\le +1$ & $\le 2$ \\
	 \ | $N$ & $0$ & $0$ & $+11$ & $+11$ & $\le 1$ \\
	 \ | $L$ & $0$ & $0$ & $+12$ & $+12$ & $\le 2$ \\
	 \ | $L'$ & $0$ & $0$ & $0$ & $0$ & $0$ \\
	 add a solid child& $\le +4$ & $\le +6$ & $0$ & $\le +10$ & $\le 3$ \\
	 \ | $G\to A$ & $\le 0$ & $\le +6$ & $0$ & $\le +6$ & $\le 2$ \\
	 \ | $A\to G$ & $\le +4$ & $\le +6$ & $0$ & $\le +10$ & $\le 3$ \\
	 \ | $N$ or $L*$ & $0$ & $0$ & $0$ & $0$ & $0$ \\
	 child removal& $\le +4$ & $\le +1$ & $\le +12$ & $\le +12$ & $\le 2$ \\
	 link & $\le +8$ & $\le +7$ & $\le +12$ & $\le +22$ & $\le 6$ \\
	 \ + $h$ removal from $p$ & $\le +4$ & $\le +1$ & $\le +12$ & $\le +12$ & $\le 2$ \\
	 \ + add $h$ as child to $s$ & $\le +4$ & $\le +6$ & $0$ & $\le +10$ & $\le 3$ \\
	 \ + $h$ type to $N$ & $\le 0$ & $\le 0$ & $\le 0$ & $\le 0$ & $\le 1$ \\
	 link of rank roots& $\le +4$ & $\le +6$ & $\le 0$ & $\le +10$ & $\le 4$ \\
	 \hline
	\end{tabular}
 \end{center}
 \vskip 4pt
 Here $p$ again denotes number of pointer changes not reflected in heap trees when arrays are used for caches.
 (Including the heap node list pointer cahnges).
\end{table}

Before a public worst case method is called, the $\Phi$ coordinates does not exceed the equilibrium values. 
During the method the coordinates changes $\Delta\Phi$ are maintained.
Worstcase version of \FindMin\ performs cache reductions as mentioned in the previous two sections.
Each other public method calls \FindMin\ and does not introduce new violations after the return.

Whenever we decrement size of the heap, we decrement the reference count as well, we two times remove first node $f$ of the list of heap nodes (if it exists), we make two degree reductions on $f$ and put $f$ to the end of the list. This makes the degree constraints to hold for all nodes of the heap (assuming they have held prior to the decrement).

Whenever we set violation type of a node $x$, we remove $x$ from the same rank identifying place of the original type (unless the type is $N$).
And we insert it to cache corresponding to the new type (unless the type is $N$ or we know the node is already there).

Whenever we decrement rank of a node $p$, violations should be made up to date. 
We should know if decrement is done by $\alpha$) rank child removal or $\beta$) rank child conversion to nonrank child.
In the case $p$ is rank root, violation type should be set to $G$ in case $\alpha$ 
and to its original value in case $\beta$\footnote{by the already described method}.
In the case $p$ is not rank root, it's loss is increased. 
If $p$ violation type is $N$ the loss becomes 1, we set violation type to $L*$ and subtype to $L$\prevfootnotemark. 
If the loss was 1 (violation type $L*$ and subtyle $L$), we set violation type to $L*$ and subtype to $L'$\prevfootnotemark.
Only in the case loss of $p$ was at least 2 (violation type $L*$ and subtype $L'$), we know $p$ has proper type and subtype and is in $LC$ so no update is needed.

Whenever we add a solid child $c$ to a rank root $p$, $p$ must have $A$ or $G$ violation type. Violation type should be set to the other\prevfootnotemark. When the type changed from $A$ to $G$ we should call the degree reduction on $p$ (what could create a new violation of type $A$) to finish the rank increment.
There is no sideeffect when we add a child to a rank child.

Removal of a child $c$ of parent $p$ means following: In all cases the parent pointer of $c$ would be set to \nill\ and $c$ would be removed from the
children list of $p$ and added to the list of heap tree roots.
If $c$ was a rank child, rank of $p$ is decremented\prevfootnotemark.

To link two solid nodes means comparing their keys, let node $s$ be the one with smaller key while $h$ the other. 
If $h$ had no parent, it is simply removed from its sibling list.
Otherwise removal of a child $h$ of its parent is invoked\prevfootnotemark.
Node $h$ is added as a solid (therefore as leftmost) child of $s$\prevfootnotemark\ marking $h$ rank child if the nodes had equal rank and nonrank otherwise.
If a rank child was added, rank of $s$ should be incremented and type of violation of $h$ set to $N$\prevfootnotemark.

\begin{table}
 \begin{center}
  \caption{Effect of public methods to violations}
	\label{tab:pubmeth}
	\begin{tabular}{lrrrrr}
	 Method & $\Phi_G$ & $\Phi_A$ & $\Phi_L$ & $\Phi$ & $p$\\
	 \hline
	 \hline
	 \Insert & $\le +4$ & $\le +12$ & $0$ & $\le +16$ & $\le 5$\\
	 \ + \FindMin\ phase 0 & $0$ &  $+6$ & $0$ & $+6$ & $1$\\
	 \ + \FindMin\ phase 2 & $\le +4$ & $\le +6$ & $0$ & $\le +10$ & $\le 4$\\
	 \hline
	 \FindMin & $0$ & $0$ & $0$ & $0$ & $0$ \\
	 \hline
	 \DeleteMin \\
	 \ + heap size decrement & $0$ & $\le 24$ & $0$ & $\le 24$ & $\le 7$ \\
	 \ + $\rho$ removal from heap nodes & $0$ & $0$ & $0$ & $0$ & $\le 3$ \\
	 \ + $\rho$ type to $N$ & $\le 0$ & $\le 0$ & $\le 0$ & $\le 0$ & $\le 1$ \\
	
	 \ + \FindMin\ phase 0 & $\le 12R(2n)$ & $\le 6R(n)$ & $0$ & $\le 12R(2n)$& $< 6R(2n)$ \\
   \ +	                 & $+20$ & & & $+2R(n)+20$ & $+10$\\
	 \ + \FindMin\ phase 2 & $\le 4R(n)$ & $\le 6R(n)$ & $0$ & $\le 10R(n)$& $\le 4R(n)$ \\
	 \hline
	 \Decrement & $\le +8$ & $\le +13$ & $\le +12$ & $\le +28$ & $\le 8$ \\
	 \ + child removal & $\le +4$ & $\le +1$ & $\le +12$ & $\le +12$ & $\le 2$ \\
	 \ + \FindMin\ phase 0 & $0$ &  $\le +6$ & $\le 0$ & $\le +6$ & $\le 2$\\
	 \ + \FindMin\ phase 2 & $\le +4$ & $\le +6$ & $0$ & $\le +10$ & $\le 4$\\
	 \hline
	 \Meld\ ($h_H$, $\Phi_{h_S}\leftarrow 0$) & $\le +8$ & $\le +6$ & $0$ & $\le +14$ & $\le 5$\\ 
	 \ + \FindMin\ phase 0& $+4$ & $0$ & $0$ & $+4$ & $1$\\
	 \ + \FindMin\ phase 2& $\le +4$ & $\le +6$ & $0$ & $\le +10$ & $\le 4$\\
	 \hline
	\end{tabular}
 \end{center}
 \vskip 4pt
 Here $p$ again denotes number of pointer changes not reflected in heap trees when arrays are used for caches.
 (Including the heap node list pointer cahnges). 

\DeleteMin\ requires at most $12R(2n)+12R(n)+44\le 14.4\log_2(2n)+14.4\log_2(n)+188\le 29\log_2n+203$  cache size reductions in amortized sense. It would generate at most $42R(2n)+40R(n)+153\le 50.4\log_2(2n)+48\log_2 n+645<99\log_2 n+696$ pointer change overhead.
If $\Phi^0$ be potential before and $\Phi^E$ after \DeleteMin, we should include the difference into account as well. But $\Phi^0\le 18R(n)+18$ and $\Phi^E\ge 0$. So we have worst case bound $12R(2n)+30R(n)+62$ cache reductions. It would generate at most $42R(2n)+94R(n)+207\le 50.4\log_2(2n)+112.8\log_2 n+699<153\log_2 n+750$ pointer change overhead in worst case.
\end{table}

\MakeHeap\ inicializes the heap structure.

\Insert($k$)\ creates new solid node $x$ with violation type $N$, key $k$, rank 0, no parent and no child, pointing to the heap. 
It increments the heap size and reference count in the heap information without side effects.
It adds $x$ as a new root to the list of heap tree roots and invokes \FindMin.
\Insert\ returns $x$ for further references. 

\FindMin\ traverses nodes of heap tree roots list and makes their parent pointers explicitly to \nill. It converts implicitly deffered roots to explicitly deffered and (even new) explicitly deffered to solid, the newly solid roots violation type is set to $G$. Violation type of a root which was already solid is checked to be either $A$ or $G$. If not, it is set to $A$.
This finishes phase 0.

In the worst case variant, the changes to $\Phi$ coordinates were calculated and cache size reductions are performed whenever corresponding $\Delta\Phi$ is positive and cache is nonempty, until all coordinates with positive $\Delta\Phi$ have empty caches. 
In the amortized variant $\Delta\Phi$ is not calculated at all and cache size reductions are performed until caches are empty.
This finishes phase 1.

Than \FindMin\ traverses the heap tree roots leftwise linking two neighbouring roots interlaced with steps to left in the circular list (to link the roots as even as possible).
We finish when only one tree remains. It's root points to minimum and it will be returned.
$\Delta\Phi$ are updated during phase 2 (of worst case variant) as well and reduction of cache sizes is repeated at the phase 3 what is last phase of \FindMin.

\DeleteMin\ implements only amortized variant, which has guaranted worst case time $O(\log n)$, so no maintainance of $\Delta\Phi$ 
coordinates is needed. It decrements size\prevfootnotemark\ in the heap information. Let $\rho$ be the only tree root. It updates pointer to the list of roots to point to the leftmost child of $\rho$. It removes $\rho$ from list of heap nodes and sets violoation type of $\rho$ to $N$\prevfootnotemark. At the end it calls \FindMin\ and discards $\rho$. 

\Decrement($x$, $k$) removes $x$ from its parent $p$\prevfootnotemark\ if such parent exists.
Than in all cases it updates key at node $x$ to $k$. It invokes \FindMin\ at the end.

\Meld($h_1$, $h_2$) identifies smaller heap $h_S$ and larger $h_H$ by comparing size info in the heap informations (call with a heap of size info $<0$ is invalid).
It appends list of $h_S$ nodes to start of the list of $h_H$ nodes (and sets corresponding pointer at $h_S$ to \nill). 
As position nodes of $h_S$ in the new list remain same, but the heap size at least doubles, $c_2\log_2(2n-p)$ increases by at least
$c_2>1$ so we got reserve 1 in degree bounds so we could make solid node with loss 0 of $h_S$ deffered node of $h_H$ without violating degree constraint bounds (for other nodes of $h_S$ it is even more obvious).
It stores sum of the sizes in the heap $h_H$ informations and changes size in $h_S$ to -1, that makes all $h_S$ nodes implicitly deffered. 
It appends roots of trees list of $h_S$ to the front of roots of trees list of $h_H$ (and sets them to \nill\ in $h_S$). 
The same rank identifying places and caches of $h_S$ are discarded.
So $h_S$ contains only negative size info and reference counts to allow discard after no implicit deffering is caused by $h_H$. 
Finally it invokes \FindMin\ and returns $h_H$ as a current heap.

We can see the effect of public methods on $\Phi$ coordinates and pointer overhead in the table \ref{tab:pubmeth}. 
Together with reduction of caches we got at most $28$ cache size reductions and $8$ in additional pointer overhed 
for amorized version (amortized) of other methods than \DeleteMin. 
This with cache reductions makes pointer overhead at most $92$ per such method.
With worst case version the upper bound is $48$ cache size reductions and $8$ in additional pointer overhed 
so overhead at most $152$ pointer changes (and constant time) is guaranted.

There is alternative not to calculate $\Delta\Phi$ coordinates during worst case methods and use their upper bounds instead and plan cache reductions according the upper bounds. If it's sufficient to guarantee $O(\log n)$ worstcase bounds, simpler strategy is to keep caches $|GC|$, $|AC|$ empty and do two $|LC|$ cache reductions after each $\Decrement$.

\section{Simplification when \Meld\ is not needed}
There will be no need for pointers to heap, no need to maintain heap size and the heap reference count. 
As deffered nodes are created only by \Meld\ method, there will be no deffered nodes in the heap at all.
Therefore all nonrank nodes will be rank roots. 
Their number is limited by their maintenance in violation lists by $2R(n)+2$.  
This makes degrees bounded by $O(\log n)$ and the degree reduction is impossible and it is not needed at all. 
All nodes have implicitly degree reserve,
so there is no need to maintain rank roots with two different violation types and one violation type say $A$ suffices.
The global node list to organize degree reductions is not needed as well.
So the only support needed are the two volation types $A$ and $L$ with same rank identifying places and caches.
If there are no deffered nodes, we would prefere inserts of nonrank nodes rather to right end of children lists for aesthetic reasons.

\begin{table}
 \begin{center}
  \caption{Effect of different transformations $|AR|+2|AC|+3|LR|+4|LC|=\Phi$}
	\label{tab:effNoDeffered}
	\begin{tabular}{lrrrrrr}
	 Reduction & $|AR|$ & $|AC|$ & $|LR|$ & $|LC|$ & $\Phi$ & $p$\\
	 \hline
	 \hline
	 $|AC|$ type $\not=A$& $0$ & $-1$ & $0$ & $0$ & $-2$ & $0$ \\
	 \hline
	 $|AC|$ type $A$ no match& $+1$ & $-1$ & $0$ & $0$ & $-1$ & $1$ \\
	 \hline
	 $|AC|$ type $A$ matched& $-1$ & $0$ & $0$ & $0$ & $\le -1$ & $2$\\
	 \hline
	 $|LC|$ type $\not=L*$ & $0$ & $0$ & $0$ & $-1$& $-4$ & $0$ \\
	 \hline
	 $|LC|$ subtype $L'$ & $\le 0$ & $\le +2$ & $\le 0$ & $\le 0$ & $\le -1$ & $\le 3$\\
	 \ - parent A in $AR$& $-1$ & $+2$ & $0$ & $\le -2$ & $\le -5$ & $3$ \\
	 \ - parent A in $AC$& $0$ & $+1$ & $0$ & $\le -2$ & $\le -6$ & $1$ \\
	 \ - parent L in $LR$& $0$ & $+1$ & $-1$ & $\le 0$ & $\le -1$ & $3$ \\
	 \ - parent L* in $LC$& $0$ & $+1$ & $0$ & $\le -1$ & $\le -2$ & $1$ \\
	 \ - parent N& $0$ & $+1$ & $0$ & $\le -1$ & $\le -2$ & $2$ \\
	 \hline
	 $|LC|$ subtype $L$ no match & $0$ & $0$ & $+1$ & $-1$& $-1$ & $1$ \\
	 \hline
	 $|LC|$ subtype $L$ matched& $\le 0$ & $\le +1$ & $\le 0$ & $\le +1$ & $\le -2$ & $\le 3$ \\
	 \ - parent of $h$ A in $AR$& $-1$ & $+1$ & $-1$ & $-1$ & $-6$ & $3$ \\
	 \ - parent of $h$ A in $AC$& $0$ & $0$ & $-1$ & $-1$ & $-7$ & $2$ \\
	 \ - parent of $h$ L in $LR$& $0$ & $0$ &$-2$ & $+1$ & $-2$ & $3$ \\
	 \ - parent of $h$ L* in $LC$& $0$ & $0$ & $-1$ & $0$ & $-3$ & $1$ \\
	 \ - parent of $h$ N& $0$ & $0$ & $-1$ & $0$ & $-3$ & $2$ \\
	 \hline
	\end{tabular}
 \end{center}
 \vskip 4pt
 Here $p$ denotes number of pointer changes not reflected in heap trees during reduction when arrays are used for caches.
 We can see each cache reduction decrements $\Phi$ by at least 1.
\end{table}

The violation reduction steps would simplify as shown in Table \ref{tab:effNoDeffered}. $\Phi$ simplifies to $|AR|+2|AC|+3|LR|+4|LC|$ with coordinates $\Phi_A=|AR|+2|AC|$ and $\Phi_L=3|LR|+4|LC|$. $\Delta\Phi$ could pay for violation reductions in amortized case, the analysis for worstcase case would show that there could be at most $\Delta^1 \Phi_A-\Delta^E \Phi_A+\Delta^1 \Phi_L+3\le 
\max(-1,\Delta^0 \Phi_A)+\max(-3,\Delta^0 \Phi_L)+4$ reduction steps in total.
Maximal degree would be $2R(n)+1$.

\begin{table}
 \begin{center}
  \caption{Effect of private blocks to violations}
	\label{tab:privblockNoMeld}
	\begin{tabular}{lrrrr}
	 Method & $\Phi_A$ & $\Phi_L$ & $\Phi$ & $p$\\
	 \hline
	 \hline
	 set violation type $A$& $\le +2$ & $\le 0$ & $\le +2$ & $\le 2$ \\
	 \ | from $A$ & $\le +1$ & $0$ & $\le +1$ & $\le 2$ \\
	 \ | from $L*$ & $+2$ & $\le 0$ & $\le +2$ & $\le 2$ \\
	 \ | from $N$ &  $+2$ & $0$ & $+2$ & $1$ \\
	 set violation type $L*$&  $\le 0$ & $\le +5$ & $\le +5$ & $\le 2$ \\
	 \ | from $A$ &  $\le 0$ & $+4$ & $\le +4$ & $\le 2$ \\
	 \ | from $L*$ & $0$ & $\le +5$ & $\le +5$ & $\le 2$ \\
	 \ | from $N$ &  $0$ & $+4$ & $+4$ & $1$ \\
	 set violation type $N$ & $\le 0$ & $\le 0$ & $\le 0$ & $\le 1$ \\
	 \ | from $A$ &  $\le 0$ & $0$ & $\le 0$ & $\le 1$ \\
	 \ | from $L*$ & $0$ & $\le 0$ & $\le 0$ & $\le 1$ \\
	 \ | from $N$ & $0$ & $0$ & $0$ & $0$ \\
	 rank decrement& $\le +1$ & $\le +5$ & $\le +5$ & $\le 2$ \\
	 \ | $A$ & $\le +1$ & $0$ & $\le +1$ & $\le 2$ \\
	 \ | $N$ & $0$ & $+4$ & $+4$ & $\le 1$ \\
	 \ | $L$ & $0$ & $\le +5$ & $\le +5$ & $\le 2$ \\
	 \ | $L'$ & $0$ & $0$ & $0$ & $0$ \\
	 add a solid child& $\le +1$ & $0$ & $\le +1$ & $\le 2$ \\
	 \ | $A$ & $\le +1$ & $0$ & $\le +1$ & $\le 2$ \\
	 \ | $N$ or $L*$ & $0$ & $0$ & $0$ & $0$ \\
	 child removal& $\le +1$ & $\le +5$ & $\le +5$ & $\le 2$ \\
	 link & $\le +2$ & $\le +5$ & $\le +6$ & $\le 5$ \\
	 \ + $h$ removal from $p$ & $\le +1$ & $\le +5$ & $\le +5$ & $\le 2$ \\
	 \ + add $h$ as child to $s$ & $\le +1$ & $0$ & $\le +1$ & $\le 2$ \\
	 \ + $h$ type to $N$ & $\le 0$ & $\le 0$ & $\le 0$ & $\le 1$ \\
	 link of rank roots& $\le +1$ & $\le 0$ & $\le +1$ & $\le 3$ \\
	 \hline
	\end{tabular}
 \end{center}
 \vskip 4pt
 Here $p$ again denotes number of pointer changes not reflected in heap trees when arrays are used for caches.
 (Including the heap node list pointer cahnges).
\end{table}

\begin{table}
 \begin{center}
  \caption{Effect of public methods to violations}
	\label{tab:pubmethNoMeld}
	\begin{tabular}{lrrrr}
	 Method & $\Phi_A$ & $\Phi_L$ & $\Phi$ & $p$\\
	 \hline
	 \hline
	 \Insert & $\le +3$ & $0$ & $\le +3$ & $\le 3$\\
	 \ + \FindMin\ phase 0 &  $+2$ & $0$ & $+2$ & $1$\\
	 \ + \FindMin\ phase 2 & $\le +1$ & $0$ & $\le +1$ & $\le 2$\\
	 \hline
	 \FindMin & $0$ & $0$ & $0$ & $0$ \\
	 \hline
	 \DeleteMin \\
	 \ + $\rho$ type to $N$ & $\le 0$ & $\le 0$ & $\le 0$ & $\le 1$ \\
	 \ + \FindMin\ phase 0 & $\le 2R(n)$ & $0$ & $\le 2R(n)$& $\le R(n)$ \\
   \ + \FindMin\ phase 2 & $0$ & $0$ & $0$& $0$ \\
	 \hline
	 \Decrement & $\le +3$ & $\le +5$ & $\le +8$ & $\le 5$ \\
	 \ + child removal & $\le +1$ & $\le +5$ & $\le +5$ & $\le 2$ \\
	 \ + \FindMin\ phase 0 & $\le +2$ & $0$ & $\le +2$ & $\le 1$\\
	 \ + \FindMin\ phase 2 & $\le +1$ & $0$ & $\le +1$ & $\le 2$\\
	 \hline
	\end{tabular}
 \end{center}
 \vskip 4pt
 Here $p$ again denotes number of pointer changes not reflected in heap trees when arrays are used for caches.
 (Including the heap node list pointer cahnges). 

 \DeleteMin\ requires at most $2R(n)\le 2.4\log_2(n)+12$ cache size reductions in amortized sense. 
  It would generate at most $7R(n)+1\le 8.4\log_2 n+43$ pointer overhead.
  If $\Phi^0$ be potential before and $\Phi^E$ after \DeleteMin, we should include the difference into account as well. But $\Phi^0\le 4R(n)+4$ and $\Phi^E\ge 0$. So we have worst case bound $6R(n)+4$ cache reductions. It would generate at most $19R(n)+13\le 22.8\log_2 n+127$ pointer change overhead in worst case.
	
\end{table}

We can see the effect of public methods on $\Phi$ coordinates and pointer overhead in the table \ref{tab:pubmethNoMeld}. 
Together with reduction of caches we got at most $8$ cache size reductions and $5$ in additional pointer overhed 
for amorized version (amortized) of other methods than \DeleteMin. 
This with cache reductions makes pointer overhead at most $29$ per such method.
With worst case version the upper bound is $12$ cache size reductions and $5$ in additional pointer overhed 
so overhead at most $41$ pointer changes (and constant time) is guaranted.

\section{Concluding remarks}
We have not discussed problems with ids to make heap keys unique.
Usually incrementing global count (much bigger range than available memory) would be sufficient.
Garbage collection could help not to increment the count too often when the heap size is maintained almost constant.
With usage where heap size oscilates among small and big sizes the garbage should be discarded to keep structure size proportional to represented set.
In such a scenario when pool of counts is going to be exhausted, nodes of all the heaps could be traversed and ordered temporary set of used id's constructed. The ids could be replaced by their order in the set. This overhead could be distributed among long enough sequence of operations. 

\section{Summary}
We have shown a variant of worst case heaps not losing information by repeated linking of heap nodes under the heap roots could be implemented and the overhead af the heaps could be kept in reasonable bounds. Especially for heaps not requiring \Meld\ operation the overhead of the heaps is small.
The overhead is probably smaller than in Fibonacci heaps as we do not discard information from same rank identifying places at the end of \FindMin.

For the worst case interface of \Decrement\ heaps (without \Meld) these are the fastest and simplest published heaps so far (according to our current knowledge).


\bibliography{ffhwce}
\end{document}